\documentclass[a4paper,12pt]{article}
\usepackage[centertags]{amsmath}
\usepackage{amsfonts}
\numberwithin{equation}{section}
\begin{document}
\title{Non--commutative phase and the unitarization of
$GL_{p,q}\left(2\right)$}

\author{M. Ar\i k and B.T. Kaynak \\
{\small Department of Physics, Bo\u{g}azi\c{c}i University, 80815 Bebek, Istanbul, Turkey}}
\date{}
\maketitle

\begin{abstract}
In this paper, imposing hermitian conjugate relations on
 the two--parameter deformed quantum group $GL_{p,q}\left(2\right)$ is studied.
This results in a non-commutative phase associated with the unitarization of the
 quantum group. After the achievement of the quantum group $U_{p,q} \left( 2 \right)$
 with $pq$ real via a non--commutative phase, the representation of the algebra
 is built by means of the action of the operators constituting the
 $U_{p,q} \left( 2 \right)$ matrix on states.
\end{abstract}

\section{Introduction}\label{int}
The mathematical construction of a quantum group $G_{q}$ pertaining
 to a given Lie group $G$ is simply a deformation of a commutative
 Poisson-Hopf algebra defined over $G$. The structure of a deformation
 is not only a Hopf algebra but characteristically a non-commutative
 algebra as well. The notion of quantum groups in physics is widely
 known to be the generalization of the symmetry properties of both
 classical Lie groups and Lie algebras, where two different mathematical
 blocks, namely deformation and co--multiplication, are simultaneously
 imposed either on the related Lie group or on the related Lie algebra.

A quantum group is defined algebraically as a quasi--triangular Hopf algebra.
 It can be either non-commutative or commutative. It is fundamentally a
 bi--algebra with an antipode so as to consist of either the q--deformed
 universal enveloping algebra of the classical Lie algebra or its dual,
 called the matrix quantum group, which can be understood as the q--analog
 of a classical matrix group~\cite{Bie}. One needs four axioms, namely morphisms,
 in order to define a bi-algebraic structure: associativity, co--associativity, unit
 and co--unit. There is an additional structure, called the connecting axiom,
 which is needed to link the algebra to its dual one.

Although the applications of quantum groups mainly concentrate on the studies of
 quantum integrable models using the quantum inverse scattering method and
 non--commutative geometry, there have been many phenomenological applications of
 quantum algebras in nuclear physics, condensed matter physics, molecular physics,
 quantum optics and elementary particle physics. The most important and remarkable
 application arose from the q--deformation of the known quantum mechanical harmonic
 oscillator algebra. The algebraic approaches to the oscillator algebras involve the
 known creation, annihilation and number operators. It is worth emphasizing the
 importance of the algebra possessing hermitian operators, giving rise to the ability
 of representing the physical observables. An algebra, therefore, needs to have a
 $\ast$ structure to be interpreted as an algebra of observables.
The simplest matrix quantum group with such a structure is $SU_q (2)$~\cite{Wor}.

Let us now review the quantum group $GL_{p,q}\left(2\right)$ and then its unitary form $U_{q,\bar q}
 \left( gl \left( 2 \right) \right)$~\cite{Jag}.
The representation matrix and the algebra of the two--parameter
 deformed quantum group $GL_{p,q}\left(2\right)$ are defined in
 the following way~\cite{Zum,Dob}
\begin{equation}
A = \left(
\begin{array}{cc}
a & b\\
c & d
\end{array}
\right) \, ,
\end{equation}
\[
\begin{array}{cccc}
ab=q^2ba & ac=p^2ca & bd=p^2db & cd=q^2dc
\end{array}
\]
\begin{equation}\label{commpq}
\begin{array}{cc}
ad-da=\left(q^2-p^{-2}\right)bc & bc = p^2 q^{-2} cb \, .
\end{array}
\end{equation}
The co--multiplication $\Delta$, the co--unit $\varepsilon$,
 and the antipode( matrix inverse) $S$,
 whose bi--algebra is generated by the matrix elements $a$, $b$, $c$,
 and $d$, are given by
\begin{equation}
\Delta \left( A \right) = \left(
\begin{array}{cc}
  a \otimes a + b \otimes c & a \otimes b + b \otimes d \\
  c \otimes a + d \otimes c & c \otimes b + d \otimes d
\end{array}
\right) \, ,
\end{equation}
\begin{equation}
\varepsilon \left( A \right) = \left(
\begin{array}{cc}
 1 & 0 \\
 0 & 1
\end{array}
\right) \, ,
\end{equation}
\begin{equation}
S \left( A \right) =
\mathcal{D}^{-1} \left(
\begin{array}{cc}
d & - p^{-2} b\\
-p^2 c & a
\end{array}
\right) = \left(
\begin{array}{cc}
d & -q^{-2} b \\
-q^2 c & a
\end{array}
\right) \mathcal{D}^{-1} \, ,
\end{equation}
where the quantum determinant of $A$ is defined by
\begin{equation}\label{detq}
\mathcal{D} \equiv det_{p,q} A = ad - q^2 bc = ad - p^2 cb \, .
\end{equation}
The co--product and the antipode of the quantum determinant are given by
\begin{eqnarray}
\Delta \left( \mathcal{D} \right) &=& \mathcal{D} \otimes \mathcal{D} \label{DD} \, , \\
S \left(\mathcal{D} \right) &=& \mathcal{D}^{-1} \label{SD} \, .
\end{eqnarray}

A unitarized form of $GL_{p,q}\left( 2 \right)$, named $U_{\bar q,q}\left(2\right)$,
 can be found in Jagannathan and Van Der Jeugt~\cite{Jag}.
It is important to notice that our notation is different from the usual one as regards the usage of the deformation parameters $p$ and $q$.
The deformation parameters $p$ and $q$ should be replaced by $p^{1/2}$ and $q^{1/2}$ to obtain the usual convention in~\cite{Jag}.
The fundamental $A$--matrix of the quantum group is given by
\begin{equation}\label{uqbq}
A = \left(
\begin{array}{cc}
a & b \\
c & d
\end{array}
\right) = \left(
\begin{array}{cc}
a & - \bar q \mathcal{D} c^\ast \\
c & \mathcal{D} a^\ast
\end{array} \right) = \left(
\begin{array}{cc}
a & -q c^\ast \mathcal{D} \\
c & a^\ast \mathcal{D}
\end{array}
\right) \, ,
\end{equation}
where the matrix elements satisfy
\[
\begin{array}{ccccc}
a c = \bar q c a & a \mathcal{D} = \mathcal{D} a & a c^\ast = q c^\ast a &
\mathcal{D} c^\ast = e^{2 i \theta } c^\ast \mathcal{D} & c c^\ast = c^\ast
c \, ,
\end{array}
\]
\begin{equation}
\begin{array}{ccc}
\mathcal{D}^\ast \mathcal{D} = \mathcal{D} \mathcal{D}^\ast = 1 & a a^\ast + \left| q
\right|^2 c^\ast c = 1 & a^\ast a + c^\ast c = 1 \, .
\end{array}
\end{equation}
Here $q = \left| q \right| e^{i \theta}$, $p = \bar q$  and
 $\theta$ is a phase. The case $\mathcal{D} = 1$ which also implies
 $\theta = 0$ corresponds to $SU_q \left( 2 \right)$.

In this paper an algebra obtained by imposing $\ast$ relations on the operators $a,b,c$ and $d$ which are the matrix elements
 of the quantum group $GL_{p,q}\left(2\right)$ will be considered.
We are able to do this for $pq$ real. In the limit $p = \bar q$, our algebra coincides with $U_{q,\bar q}\left( 2 \right)$~\cite{Jag}.
We thus name this algebra $U_{p,q}\left(2\right)$. Representation of this algebra is constructed and the relationship of
 these representations to q--oscillators and to two--parameter coherent states are discussed.
\section{The Unitarization of the Quantum Group $GL_{p,q}
 \left( 2 \right)$ with $p \ne \bar q$}
 In order to obtain $SU_q \left( 2 \right)$, elements of
  the fundamental $A$--matrix, $A \in GL_q \left( 2, \mathbb{C} \right)$,
  are chosen in such a way that $A^\ast = A^{-1} = 1$.
  This choice brings about a restriction on the elements
  of the matrix such that $b = -q c^\ast$ and $d = a^\ast$.
  The procedure applied to the quantum group $GL_{p,q} \left( 2,
  \mathbb{C} \right)$ to carry out the unitarization is
  similar to the one applied to the one--parameter deformed
  quantum group in order to transform $GL_q \left( 2, \mathbb{C}
  \right)$ into $U_q \left( 2 \right)$ but it is not completely the same.
The most important point of the procedure we have studied is
  that the matrix of the quantum group should be factorized into a
  product which consists of the square root of the quantum determinant
  and a new matrix whose determinant is unity.
\begin{eqnarray}
  \mathcal{D}  &=& \delta^2 \, , \\
 A &=& \delta A_{right} \, .
\end{eqnarray}
The co--product and the antipode of $\delta$ are given by
\begin{eqnarray}
\Delta \left( \delta \right) &=& \delta \otimes \delta \, , \\
S\left( \delta \right) &=& \delta ^{-1}\, ,
\end{eqnarray}
which are consistent with the equations~(\ref{DD}) and~(\ref{SD}).
 The next step is to impose the unitarity condition on the new matrix, resulting in finding
  the relation between the elements of this matrix as in the
  $U_q \left( 2 \right)$ case. Therefore, the elements of the matrix $A_{right}$  
become the elements of $SU_r(2)$ with $r \in \mathbb{R}$ 
\begin{equation}
r = pq = \bar p \bar q \, .
\end{equation} 
Lastly, the relations between
  the original matrix elements can be achieved through the relations
  between the new ones obtained after the unitarization of the matrix with the condition
\begin{equation}
\delta \delta^\ast = s^2 \delta^\ast \delta \, ,
\end{equation}
where $s$ is a central element of the resultant unitarized algebra of $GL_{p,q} \left(2 \right)$.
It commutes with all elements in the algebra and is also hermitian $s = s^\ast$.
The co--product, co--unit and the antipode of the central element $s$ are given by
\begin{eqnarray}
\Delta \left( s \right) &=& s \otimes s \label{Ds} \, , \\
\varepsilon \left( s \right) &=& 1 \, , \\
S \left( s \right) &=& s^{-1} \label{Ss} \, .  
\end{eqnarray}
This leads to the matrix elements $b$ and $d$ being respectively replaced by
  a combination of $c^\ast$ and $a^\ast$ multiplied by inverse of $s$ and a unitary operator $u$.
 The new fundamental $A$--matrix of the unitary
  quantum group is given by
\begin{equation}\label{upq}
A = \left(
\begin{array}{cc}
a & b \\
c & d
\end{array}
\right) = \left(
\begin{array}{cc}
a & - \bar q^2 s^{-1} u c^\ast \\
c & s^{-1} u a^\ast
\end{array}
\right) \, .
\end{equation}
It can be easily checked that with these relations $a$, $c$ and $b$, $d$ defined by~(\ref{upq})
 satisfy the commutation relations~(\ref{commpq}) of $GL_{p,q}\left( 2 \right)$.
The co--product, the co--unit and the antipode of the unitary operator $u$ are given by
\begin{eqnarray}
\Delta \left(u \right) &=& u \otimes u \label{Du} \, , \\
\varepsilon \left( u \right) &=& 1 \, , \\ 
S \left( u \right) &=& u^\ast \label{Su} \, .
\end{eqnarray}
The whole algebra of the unitarized two--parameter quantum group, which the matrix elements obey, is given by
$$
\begin{array}{cc}
a c = p^2  c a & a^\ast c^\ast = \bar p ^{-2} c^\ast a^\ast 
\end{array}
$$
$$
\begin{array}{cc}
a c^\ast = q^2 s^2 c^\ast a & a^\ast c = \bar q ^{-2} s^{-2} c a^\ast 
\end{array}
$$
$$
\begin{array}{cc}
a a^\ast - s^2 a^\ast a = \left( 1 -r^2\right) c c^\ast & c c^\ast = \frac{q \bar q}{p \bar p} s^2 c^\ast c 
\end{array}
$$
$$
\begin{array}{cc}
u a = s^2 a u & u^\ast a^\ast = s^{-2} a^\ast u^\ast 
\end{array}
$$
$$
\begin{array}{cc}
u a^\ast = s^2 a^\ast u & u^\ast a = s^{-2} a u^\ast 
\end{array}
$$
$$
\begin{array}{cc}
u c = \frac{p \bar q}{\bar p q} s^2 c u & u^\ast c^\ast = \frac{p \bar q}{\bar p q} s^{-2} c^\ast u^\ast 
\end{array}
$$
$$
\begin{array}{cc}
u c^\ast = \frac{\bar p q}{p \bar q} s^2 c^\ast u & u^\ast c = \frac{\bar p q}{p \bar q} s^{-2}c u^\ast 
\end{array}
$$
\begin{equation}\label{comm}
u u^\ast = u^\ast u = 1 \, .
\end{equation}
It can be shown that the commutation relations~(\ref{comm}) satisfy the co--product algebra homomorphism
 and the antipode algebra anti--homomorphism.
\section{The Representation of the Quantum Group $U_{p,q} \left( 2
 \right)$}
The operators constituting an $SU_q \left( 2 \right)$
 matrix which corresponds to $\mathcal{D} = 1$, $q$ real in
 (\ref{uqbq}) can be represented by their action on states
$\left|n,m \right
 \rangle$ where $n$ is non--negative
integer corresponding to the particle number associated with the
 creation operator $a^\ast$ and $m$ is a positive or negative
 integer associated with the Fourier transform of $c$~\cite{Ari}.
 This serves two purposes. One is that it proves the algebra
 presented in the previous section is consistent. The second is
 that it gives physical insight on the oscillator properties of the operators.
 The action of the operators $a$, $a^\ast$ and $c$, $c^\ast$ of $SU_q \left( 2 \right)$
 on the states $\left|n,m \right \rangle$ is given by
\begin{eqnarray}
a \left| n,m \right \rangle & = & \sqrt{1-q^{2n}} \left| n - 1,
m \right \rangle \, , \label{aa} \\
a^\ast \left| n,m \right \rangle & = & \sqrt{1-q^{2n+2}} \left| n + 1,
m \right \rangle \, , \\
c \left| n,m \right \rangle & = & q^n \left| n,m - 1 \right \rangle \, , \\
c^\ast \left| n,m \right \rangle & = & q^n \left| n,m + 1 \right \rangle \label{cast} \, .
\end{eqnarray}
Here $m$ is an integer and $n$ is a nonnegative integer.
Motivated by this, we look for a representation of the algebra~(\ref{comm}) on such states.
The deformation parameters $p$ and $q$ are reparametrized in order to achieve a convenient form for the representation
\begin{equation}
\begin{array}{cc}
p = \sqrt{\frac{r}{t}} e^{i \theta / 2} & q = \sqrt{r t} e^{-i \theta / 2} 
\end{array} \, ,
\end{equation}
where $r$, $t$ and $\theta$ are real independent parameters.
The parameters $r$ and $t$ are positive by definition and $\theta$ is a phase angle.
The representation also depends on a real integer parameter $k$ associated with the eigenvalue of the central element $s$.
The special case where $t = 1$, $\theta = 0$, $k = 0$, and therefore $p = q$,
 corresponds to the $SU_q \left( 2 \right)$ algebra for which it is necessary that $q\in \left( 0,1\right)$
 whereas the case $k = 0$, $t = 1$ corresponds to $U_{q,\bar q} \left( 2 \right)$ discussed in Section~(\ref{int}).
The operators $c$, $c^\ast$, $a$, $a^\ast$, $u$, $u^\ast$ and $s$ act on states
\begin{eqnarray}
c \left| n,m \right \rangle & = & r^n \left( t e^{-i \theta} \right)^{m - 1}
 \left| n,m - \left( k + 1 \right) \right \rangle \, , \\
c^\ast \left| n,m \right \rangle & = & r^n \left( t e^{i \theta} \right)^{m + k}
 \left| n,m + \left( k + 1 \right) \right \rangle \, , \\
a \left| n,m \right \rangle & = & \sqrt{1 - r^{2n}} \left( t e^{-i \theta} \right)^m
 \left| n - 1, m - k \right \rangle \, , \\
a^\ast \left| n,m \right \rangle & = & \sqrt{1 - r^{2n + 2}}
 \left( t e^{i \theta} \right)^{m + k} \left| n + 1, m + k \right \rangle \, , \\
u \left| n ,m \right \rangle & = & e^{i \left( k -2m \right) \theta}
 \left| n, m - 2k \right \rangle \, , \\
u^\ast \left| n,m \right \rangle & = & e^{i \left( 2m +3k \right) \theta}
 \left| n, m+2k \right \rangle  \, , \\
s|n,m \rangle &=& t^k |n,m \rangle \, , 
\end{eqnarray}
which explicitly shows that $u$ is actually a non--commutative unitary phase operator.
 It can be easily seen that setting $t = 1$, $\theta = 0$, $k = 0$ leads to $p = q$.
 The representation above then reduces to the representation~(\ref{aa}--\ref{cast}) by the replacement of $p$ and $q$ by $p^{1/2}$ and $q^{1/2}$.
\section{Conclusion}
The most interesting aspect of our construction is the appearance of the non--commutative phase
 described by the unitary operator $u$. The work of ~\cite{Ari} was
 motivated by generalizing $SU_q \left( 2 \right)$, resulting in
 introducing a parameter $p$ and operator $d$ satisfying
 \begin{equation}
 d \left| n , m \right \rangle = p^{1 - m} \left| n , m - 1 \right \rangle \, .
 \end{equation}
 The action of this operator has some resemblance to the action of
 the square root of the quantum determinant $\mathcal{D}$ which can be shown to be given by
 \begin{equation}
  \delta \left| n ,m \right \rangle  = \left( t e^{-i \theta} \right)^m  \left| n, m-k \right \rangle \, .
 \end{equation}
 \cite{Ari} claimed that the operator $d$ can be interpreted as a deformation
 of a phase operator. The work of this paper shows that a rigorous foundation
 for a non--commutative unitary phase operator lies in the two--parameter
 deformed quantum group.

 Whether applications such as the quantum phase operator for a quantized boson
 can be incorporated into this formalism will be the subject of further research.

\end{document}